# Quantum two-way time transfer over a hybrid free-space and fiber link

Xiao Xiang[1,4], Bingke Shi[1,2,4], Runai Quan[1], Yuting Liu[1,2], Zhiguang Xia[1,2], Huibo Hong[1,2], Tao Liu[1,2], Jincai Wu[3], Jia Qiang[3], Jianjun Jia[3,†], Shougang Zhang[1,2,‡], Ruifang Dong[1,2,*]

[1]Key Laboratory of Time and Frequency Primary Standards, National Time Service Center, Chinese Academy of Sciences, Xi'an 710600, China

[2]School of Astronomy and Space Science, University of Chinese Academy of Sciences, Beijing, 100049, China

[3]Key Laboratory of Space Active Opto-Electronic Technology, Shanghai Institute of Technical Physics, Chinese Academy of Sciences, Shanghai, 200083, China

[4]Authors contributed equally to this work

Corresponding author: [*]dongruifang@ntsc.ac.cn, [†]jjjun10@mail.sitp.ac.cn, [‡]szhang@ntsc.ac.cn

**Abstract**

As the superiority of quantum two-way time transfer (Q-TWTT) has been proved convincingly over fiber links, its implementation on free space links becomes an urgent need for remote time transfer expanding to the transcontinental distance. In this letter, the first Q-TWTT experimental demonstration over a hybrid link of 2 km-long turbulent free space and 7 km-long field fiber is reported. Despite the significant loss of more than 25 dB and atmospheric turbulence, reliable time transfer performance lasting for overnights has been realized with time stability in terms of time deviation far below 1 picosecond. This achievement shows the good feasibility of quantum-enhanced time transfer in the space-ground integrated optical links and nicely certifies the capability of Q-TWTT in comparing and synchronizing the state-of-the-art space microwave atomic clocks.

**Introduction**

Clock synchronization has played great roles in fundamental research and engineering applications [1-4], including precise navigation and positioning, radio astronomy, and cutting-edge quantum information processing. To achieve high-precision synchronization between remote clocks, numerous time transfer technologies have been intensively developed. As the propagation of the timing signal can be independent of the actual link length provided that a symmetric path is shared, the two-way time transfer (TWTT) is preferred in practical applications for its small calibration uncertainty and long-term stability. For instance, the two-way satellite time and frequency transfer (TWSTFT) has been recognized as the primary technique for the generation of coordinated universal time (UTC), which achieved time uncertainty better than 1 ns [5] and time stability over a 24-hour period as low as 100 picoseconds in some systems [6]. By adopting optical pulses as the carrier of the timing signal, the time transfer performance over the satellite-ground link was further improved, reaching a time stability of 10 ps at 1 day and an accuracy better than 100 ps [7, 8]. With the maturity of optical fiber communication networks, fiber links have attracted the most attention for precise ground-based time transfer [9-11]. Among them, the TWTT over fiber links has achieved time stability of less than 6 ps/s and 0.9 ps/100s without active fiber noise suppression [12]. However, due to the inherent classical noise, the precisions of both the space-based and ground-based TWTT methods are limited to the picosecond level. Besides, secure time transfer has been well recognized as crucial since any malicious spoofing may lead to malfunctions of the key facilities relying on precise timing [13]. Although the TWTT has been found to be a requisite for avoiding the man-in-the-middle (MITM) delay attacks, the security loophole cannot be completely closed classically [14].

By virtue of the strong temporal correlation characteristics of energy-time entangled photon pair sources, the quantum-enhanced two-way time transfer (Q-TWTT) was recently proposed [15, 16] and demonstrated over coiled fiber links extending to a length of 50 km, reaching time stability of 54.6 fs at 57300 s and accuracy of 1.3±36.6 ps [17]. The results showed significant improvement in comparison with its classical counterparts. [12]. The subsequent field implementation of two-way quantum synchronization between remote clocks across a 7 km deployed fiber link further verified its capability in synchronizing the state-of-the-art microwave atomic clocks [18]. Recently, the feasibility of using Q-TWTT to distribute time across with satellites-equipped quantum resources was further studied with detailed numerical simulations, indicating a bright prospect of building a global synchronization network with precision up to tens of picoseconds.[19]. Furthermore, the security of the clock synchronization can be reliably guaranteed by the complementarity principle of quantum mechanics and requisite quantum techniques. For instance, the two-way quantum key distribution technique has been applied to ensure the security of the time information transferred between the Micius quantum satellite and the ground station[20].

In this letter, the Q-TWTT based on the energy-time entangled photon pair source was for the first time demonstrated over a hybrid link of 2 km open-air free-space and 7 km deployed single-mode fiber (SMF). The total loss of the free-space transmission was measured to be more than 25 dB, which is comparable to that expected on an optical link from the ground to the low Earth Orbit (LEO) satellite[21].Under slightly different system settings, three overnight measurements were carried out. From the measured results involving more than 10 consecutive hours under the optimized configuration, the time transfer stability showed a time deviation (TDEV) of 1.71 ps at an averaging time of 50 s and 144 fs at 6400 s. The result is well comparable with that achieved over fiber links, which clearly proves the advantage of Q-TWTT over turbulent free-space optical links in improving the time transfer precision and paves the way to its application in the space-ground integrated time and frequency network.

**Experimental setup and evaluation model**
An overview of the Q-TWTT experimental setup is illustrated in Fig. 1(a). The line-of-sight distance between the campus of National Time Service Center, Chinese Academy of Sciences (NTSC, with altitude ~475 m) and the Lishan Observatory

(LSO, with altitude~1000m) is about 2 km. With the help of a full-duplex bidirectional optical terminal installed at each site (OTA&OTB), the direct free-space optical link (marked with the solid green line) was successfully built to transmit photons across the turbulent air. A more detailed schematic diagram of the optical terminals along with the link is shown in Fig. 1(b), each of which is composed of a 180-mm-diameter Cassegrain telescope, a beacon laser (532/671 nm), and an acquisition, tracking and pointing (ATP) system. In the pre-alignment stage, the deflection angle of each telescope should be carefully adjusted so that the received beacon beam can enter the field of view of its rear photoelectric position sensitive detector (PSD) for producing horizontal and vertical positions error signals. By controlling the fast-steering mirror (FSM) according to the error signals, the ATP systems realize the fine alignment between the two optical terminals. Under the alignment condition, the co-propagated quantum photons with the beacon beam were spatially demultiplexed by a dichroic mirror (DM) for subsequent transmission or detection.

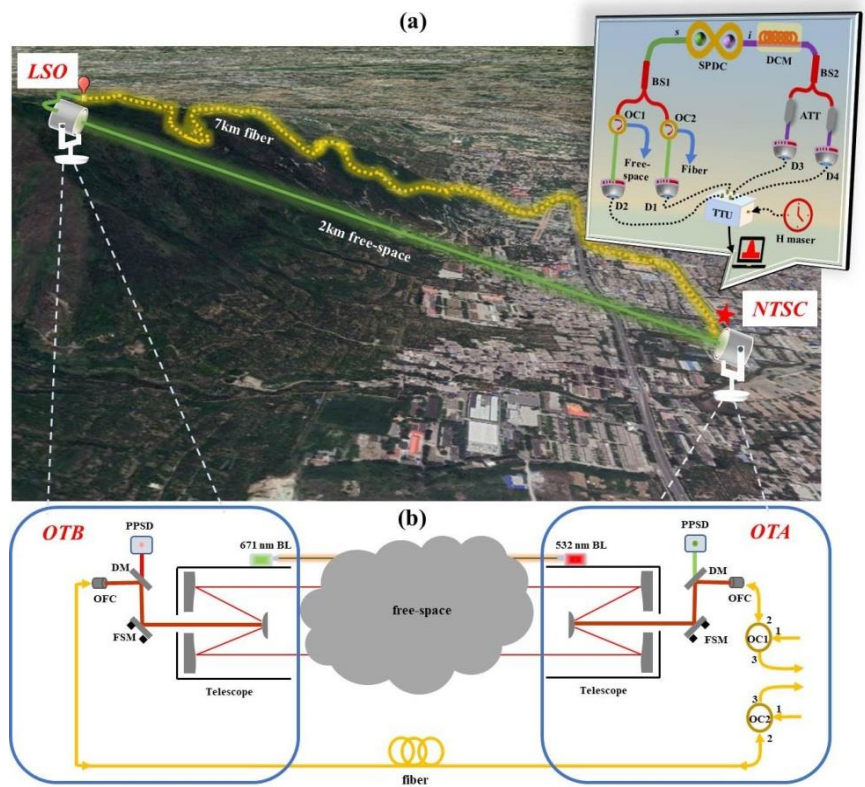

Fig. 1 Experimental setup. (a) Schematic diagram of the quantum two-way time transfer over a hybrid optical link consisting of 2 km of free space and 7 km of field fiber connecting the campus of National Time Service Center (NTSC, altitude ~475 m) and the Lishan Observatory (LSO, altitude ~1000 m); (b) The optical layout of the full-duplex bidirectional optical terminal, the Cassegrain-type telescope has a primary parabolic reflector with a diameter of 180 mm. SPDC: spontaneous parametric down-conversion biphoton source; BS1/BS2: balanced fiber beam splitters; OC1/OC2: fiber optical circulators; DCM: dispersion compensation module which was a combination of 3km-long SMF and a DCF module; ATT: optical attenuators; D1-D4: four superconducting nanowire single-photon detectors (Photec Ltd.); TTU: time tagger unit (Time

tagger ultra, Swabian Instruments); H maser: hydrogen maser with 10 MHz output. BL: beacon laser with wavelength of 532 nm/671 nm; FSM: fast-steering mirror with two piezoactuators; DM: dichroic mirror for splitting the 1560 nm photons with beacon laser; PSD: photoelectric position sensitive detector; OFC: optical-to-fiber coupler.

In addition, a 7 km-long field fiber (marked with the dashed yellow line) was also deployed between the two sites. Through the optical-to-fiber coupler (OFC) in the LSO, exchange between the free-space and single-mode fiber links can be efficiently realized. Thus, the bidirectional transmissions over the hybrid link share a common local site at NTSC, and the time transfer performance can be locally implemented, measured, and evaluated. The transmission loss of the 7 km-long field fiber has been evaluated as 2.5 dB[18], while the losses of the free-space uplink and downlink were estimated to be 23 dB and 27 dB, respectively. More detailed analysis about the effect of atmospheric turbulence on photon transmission is indicated in Appendix A.

At the NTSC campus, the self-developed all-fiber telecom-band energy-time entangled biphoton source with a photon-pair generation rate over 8 MHz/mW [22] was utilized for the Q-TWTT demonstration. The generated polarization-orthogonal photon pairs, denoted as signal and idler, were spatially departed by a fiber polarization beam splitters (FPBS) into two fiber paths respectively. Subsequently, the signal photons were split into two beams by a 50/50 fiber-based beam splitter (BS1), with each one followed by a fiber optical circulator (OC1&OC2). Via the OFCs, the second-port output from OC1 was coupled to the optical terminal for the free-space uplink transmission to LSO and then propagated back toward NTSC via the fiber link. Meanwhile, the second-port output from OC2 was directly transmitted through the fiber to LSO and then back from the free-space downlink transmission. Afterwards, the round-trip signal photons were collected from the third-port outputs of the OCs. As the loop-back optical link contains a 7 km-long field fiber, which is the main source of dispersion and will broaden the coincidence width significantly, appropriate dispersion compensation is necessary. For the idler photons, a combination of 3 km-long SMF and a 1.25 km-long dispersion compensation fiber (DCF, nominal compensating for 10 km SMF) was inserted for nonlocally compensating the dispersion introduced by the 7 km-long deployed fiber link. Afterwards, another 50/50 fiber-based beam splitter (BS2) was used to split the idler photons into two counterparts. To avoid saturating the photon detectors in long-period operation, 13 dB of attenuation was applied to limit the idler photon countrate to the level of ~320 kHz. Finally, the signal and idler photons were detected by four superconducting nanowire single-photon detectors (SNSPDs, Photec Ltd., D1-D4) and their times of arrival were recorded with a time tagger unit (TTU, Time tagger ultra, Swabian Instruments) which was referred to an H maser.

With the recorded four series of timestamps, respectively identified as $t_1, t_2, t_3, t_4$, the time offset ($t_0$) can be deduced by subtracting the bidirectional transmission delay, i.e. $t_0 = \frac{(t_2 - t_4) - (t_1 - t_3)}{2}$. Statistically, the standard deviation (SD) of $t_0$ is determined

by

$$\Delta t_0 = \frac{1}{2}\sqrt{\Delta^2(t_1 - t_3) + \Delta^2(t_2 - t_4)}, \qquad (1)$$

where $\Delta(t_1 - t_3)$ and $\Delta(t_2 - t_4)$ represent the SDs of the measured time delays after propagation through the free-space uplink and downlink, respectively. According to the quantum theoretical model, $\Delta(t_1 - t_3)$ and $\Delta(t_2 - t_4)$ can be determined by the detected two-photon temporal coincidence distributions with $1/e$ widths ($\sigma_u$ for the uplink and $\sigma_d$ for the downlink) and the photon pair numbers ($N_u$ for the uplink and $N_d$ for the downlink) within a certain measurement time [16]. Thus, the SD of $t_0$ is determined by[16]:

$$\langle \Delta t_0 \rangle_N = \frac{1}{2}\sqrt{(\frac{\sigma_u}{\sqrt{2N_u}})^2 + (\frac{\sigma_d}{\sqrt{2N_d}})^2}. \qquad (2)$$

It is clear to see that, the time offset measurement precision can be improved by narrowing the detected two-photon coincidence widths and increasing the detected photon pair numbers. Under the practical circumstance, due to the unavoidable dark counts of the SNSPDs as well as contributions from the background noise such as stray light, the accidental coincidences are also involved in the detected photon pair numbers and play a deteriorate role in the measurement precision. Thus the coincidence-to-accidental ratio (CAR), [23] which is defined as the ratio of correlated events to the system noise, should be introduced and the above expression is reformed as

$$\langle \Delta t_0 \rangle_N = \frac{1}{2}\sqrt{(\frac{\sigma_u}{\sqrt{2N_u/(1+1/CAR_{12})}})^2 + (\frac{\sigma_d}{\sqrt{2N_d/(1+1/CAR_{34})}})^2}, \qquad (3)$$

where $CAR_{13}$ and $CAR_{24}$ respectively denote the quantified coincidence-to-accidental ratios depending on the coincidence measurements of D1-D3 and D2-D4.

**Results and Analysis**

Without loss of generality, the experimental results taken during the Modified Julian Date (MJD) of 59808 to 59814 were presented. To avoid the background noise to the single-photon detection arising from the solar radiation, the experiments were performed during nights and early mornings. With the addition of a 3 km-long SMF and a DCF module, into the idler photon paths for nonlocal dispersion cancellation, the resultant coincidence widths in full width at half maximum (FWHM) were measured to be ~157 ps and ~170 ps for the uplink and downlink respectively. In this case, typical histograms of two-photon coincidences accumulated within the duration of 50 seconds are shown in Fig. 2 (a)(b). To further eliminate the residual fiber dispersion, a newly developed method for optimizing nonlocal dispersion cancellation by wavelength tuning[24] was applied, leading to narrowed widths of ~120 ps in the subsequent measurements, which was shown in Fig. 2(c)-(f). By further improving the alignment of telescopes in the optical terminal, which ensured an optimized free-space transmission loss, more coincidence counts were acquired as shown in Fig. 2 (e)(f). One may note that the histograms in Fig. 2(b)(d)(f) show higher coincidence count peak and lower noise floor than those in Fig. 2(a)(c)(e), presenting unbalanced background noise between the uplink and downlink. [25]

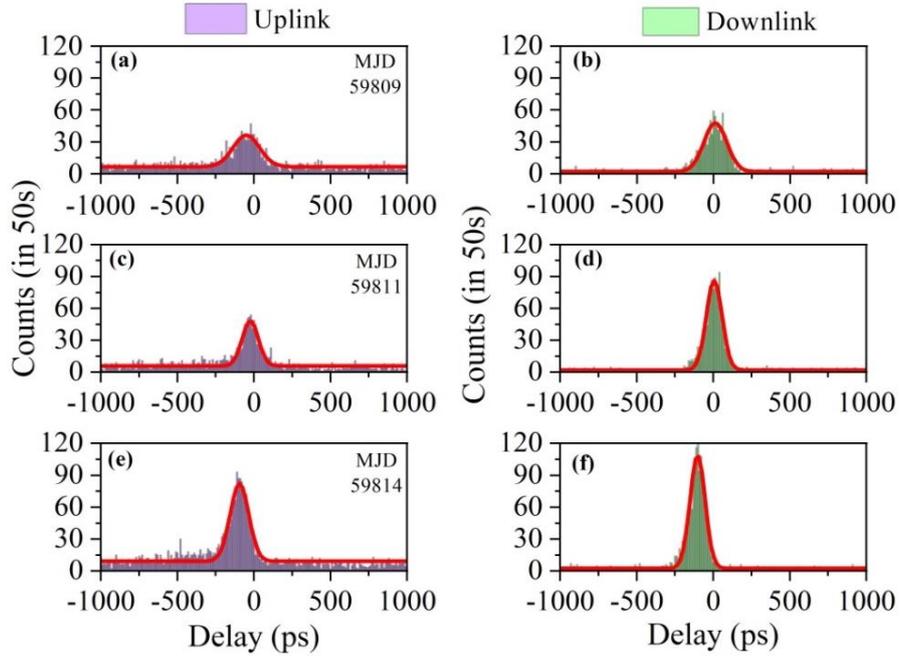

Fig. 2 Typical histograms of two-photon coincidences after the uplink (a, c, e) and downlink (b, d, f) transmissions, which were measured in each beginning of the MJD 59809, MJD 59811 and MJD 59814, respectively. Peak centers have been moved around the zero-delay position for ease of comparison.

The statistical results of coincidence widths along with coincidence counts within a measurement time of 50 seconds are displayed in Fig. 3(a)(b). As a quantitative indicator of the effective photon pairs compared to the systematic noise, the CARs under the above three situations were also measured and presented in Fig. 3(c). It is shown that, there remains an almost 3-fold unbalance of the CAR between the uplink and downlink even though the two-photon coincidence measurement is optimized both in dispersion compensation and link loss. Such unbalance should be attributed to the performance difference between the two optical terminals and variations in the strength of the optical turbulence along the slant propagation path.[25] According to Eqn. (2), the expected time offset deviations ($\langle \Delta t_0 \rangle_N$) are calculated and shown in Fig. 3 (d), indicting an improved time deviation of one picosecond level could be reached by narrowing the detected two-photon coincidence widths and increasing the detected photon pair numbers.

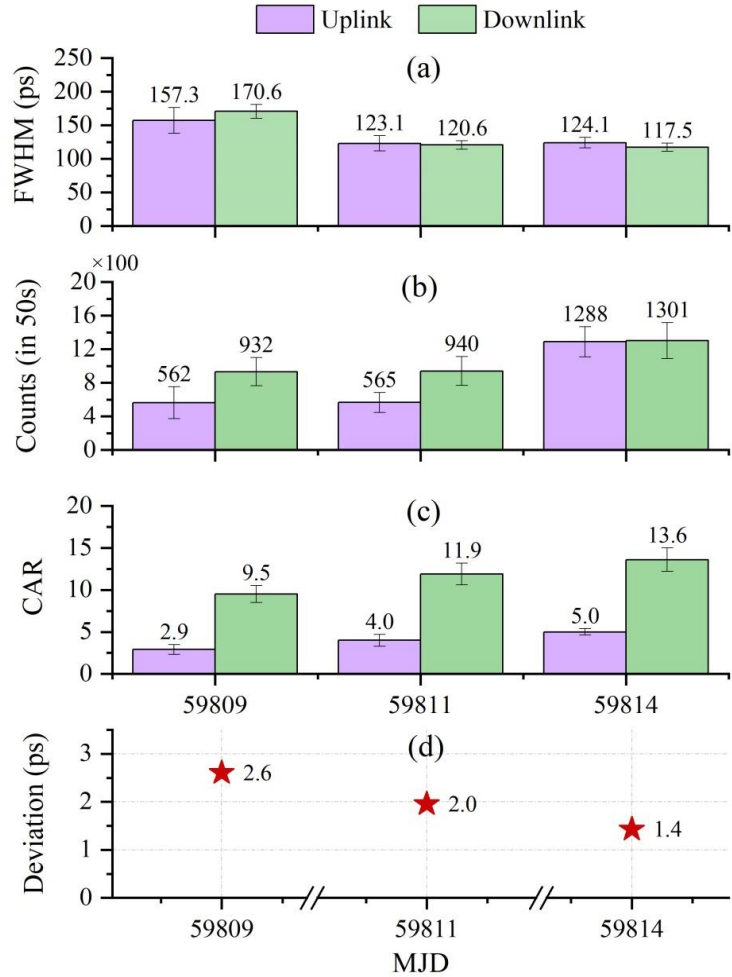

Fig. 3 Collections of coincidence widths, coincidence counts, CARs and corresponding time offset deviations extracted from the measured of two-photon coincidence histograms.

Taking the single run of measurement time as 50 seconds, the time delays of the uplink and downlink can be extracted by Gaussian fitting of the temporal centers from the measured coincidence count distributions during each run, which are plotted in Fig. 4(a) with black solid and red dashed lines respectively. It can be observed that the up- and down- link have similar time delay fluctuations, which could reach up to 500 ps over a 10-hour period. To investigate its ambient dependence, the outdoor temperature of MJD 59814 was also recorded and depicted in Fig. 4(a) with a green solid line, showing a high degree of consistency with time delay fluctuations. After deducting symmetric delay in the bidirectional scheme, the time offset results were obtained and shown in Fig. 4(b). By comparing the measured standard deviations (SDs) during the MJD 59809, MJD 59811 and MJD 59814 (which are 3.8 ps, 2.2 ps and 1.6 ps, respectively) with the theoretical simulations given in Fig. 3(d), nice agreements have been achieved. Therefore, just as predicted by the theoretical analysis, narrower coincidence width, better CAR and more photon pairs would lead to a smaller SD.

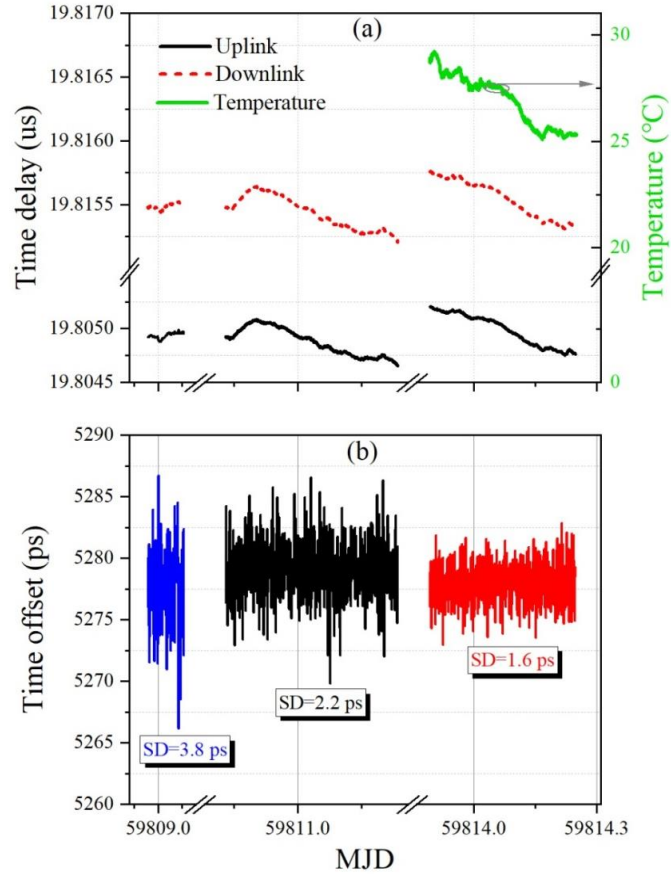

Fig. 4 Measured time delay in the uplink ($t_2 - t_4$) and downlink ($t_1 - t_3$) (a) and recovered time offset ($t_0$) by subtracting the bidirectional transmission delay (b) from MJD 59808 to MJD 59814.

The time stability performances during the measurement periods of MJD 59809, MJD 59811 and MJD 59814 were then evaluated in terms of time deviation (TDEV) [26] and presented in Fig. 5 with different colors for comparison. Despite apparent discrepancy at the initial averaging time of 50 s, the three curves all follow the slope of $\tau^{-1/2}$ until the averaging time of 1000 s, indicating the dominant contribution of white phase noise in the overall system. For the best case of optimized dispersion cancellation and free-space channel loss, the TDEV of $t_0$ was 1.71 ps at averaging time of 50 s and reached 144 fs at 6400 s (as shown in Fig. 5 by red squares and line). In addition, the TDEV of the state-of-the-art space atomic microwave clock [27] was also depicted in Fig. 5 with a dashed gray line. It can be seen that, the residual instability of the time transfer via such link is better than that of current space microwave atomic clock after the averaging time of 10 s, which can well satisfy the requirement for time synchronizing between space microwave atomic clocks.

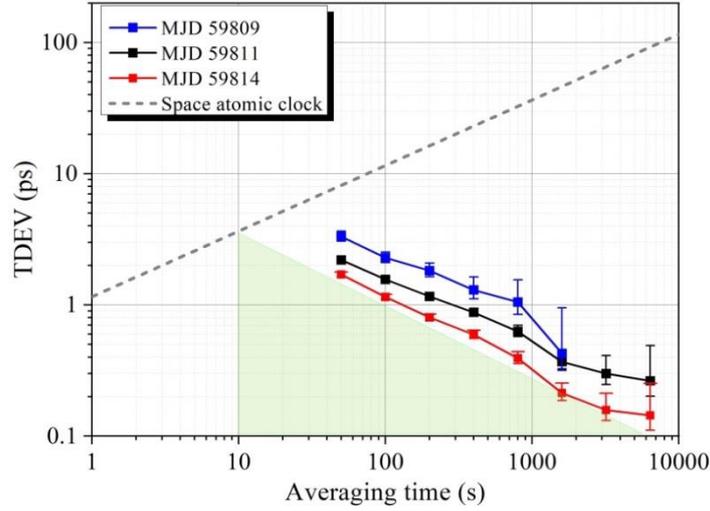

Fig. 5 Measured TDEVs of recovered time offset from MJD 59809 to MJD 59814 (blue, black and red squares and lines). The dashed gray line represents the current space microwave atomic clock performance[27], which is calculated from modified Allan deviation (MDEV) [28] (see Appendix B for detail).

In the current demonstration, the total one-way loss of the 2 km free-space link (~25dB) is comparable to that expected on an optical link from the ground station to the low Earth Orbit (LEO) satellite. According to the reported literature [29, 30], the turbulence across ~2 km-long slant atmospheric channel could be equivalent to that in the satellite-ground free space link. Therefore, the presented achievement should be applicable to ground-to-space time transfer by enlarging the optical terminal aperture and improving the photon pair source brightness. It is worth mentioning that, the time and frequency transfer scheme based on femtosecond frequency combs and phase-locking techniques is another strategy that can satisfy the comparison requirements between state-of-the-art optical clocks [31]. However, its time transfer performance is strongly dependent on the achievable noise correction bandwidth that experiences reciprocal reduction with the distances involved. [29] By contrast, the presented Q-TWTT method utilizes the bidirectional link to passively eliminate the symmetric delay fluctuation, thus there is no loop bandwidth limit when extending the transmission distance.

With the development of free-space quantum communication, a variety of strategies [32-35] have been adopted to achieve quantum key distribution (QKD) under daylight conditions by circumventing the high solar background noise. In particular, a recent literature reported a continuous QKD manipulation over the course of three and a half days across an urban free-space optical link [36]. Thus, incorporating the newly developed technologies, Q-TWTT is highly prospective to obtain equivalent performance under different lighting and weather conditions, and replace conventional T2L2 approaches for high-precision applications. Moreover, in the Q-TWTT method, due to the fundamentally random timing of the photon pair production and highly temporal correlation between the paired photons, the timing signals carried by the correlated photon pairs can be regarded as authenticated by nature. Together with the line-of-sight free space link and bidirectional transmission to avoid the MITM delay attacks, the Q-TWTT in free space channel provides an ideally secure time transfer strategy.

## Conclusion

In conclusion, the Q-TWTT based on energy-time entangled photon pair sources has been for the first time demonstrated over a 2 km-long turbulent free-space optical link, which incorporates a 7 km-long field fiber as well. Long-term operation reliability of the system lasting for overnights has been verified, and sub-picosecond time stability has been achieved, which presents its superior capability of synchronization between state-of-the-art space microwave atomic clocks. This achievement well verifies the potential enhancement of quantum time transfer in space-to-ground and interspace optical links, and paves the way for applying quantum technologies in the space-ground integrated time and frequency network.

## Appendix A

To reveal the effect of atmospheric turbulence on photon transmission, the photon countrates received in the LSO site via one-way free-space and fiber links were recorded and shown in Fig. 6. Compared with the stable countrate in the fiber link (blue line), frequent signal fading was observed in the free-space link (red line) from Fig. 6(a). To be more specific, during the 10 seconds measurement time, 336 data points with zero-count occurred in the detected 10000 samplings in the free-space link, which can be attributed to strong atmospheric turbulence caused beam scintillation, beam wander and angle-of-arrival jitter.[30] The power spectral density (PSD) of the two countrate traces is shown in Fig. 6(b), indicating the atmospheric induced fluctuations is 15dB more than that in the fiber link in the low frequency region around 1 Hz. In addition, PSD in the frequency region below 20 Hz follows a power-law distribution of $f^{-2/3}$, which is consistent with the angle of arrival spectrum predicted in the Kolmogorov turbulence model.[29]

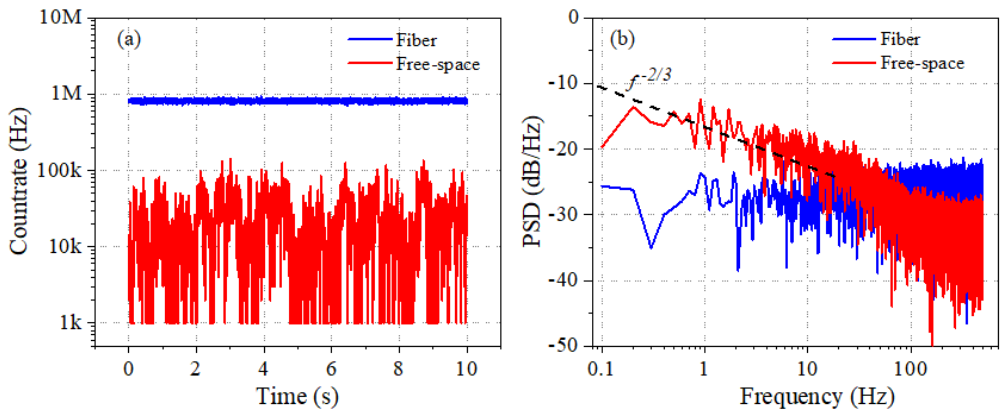

Fig. 6 Comparison of countrates received in LSO site via free-space (red) and fiber (blue). (a) Countrates versus time over 10 seconds. (b) Power spectral density (PSD) of the two countrate traces.

**Appendix B**

The time Allan deviation (TDEV) is regarded as a useful tool for characterizing the stability of a time source or distribution system, which is defined based on the modified Allan variance (MDEV): [26]

$$TDEV(\tau) = \left(\frac{\tau}{\sqrt{3}}\right) \cdot MDEV(\tau). \tag{4}$$

MDEV can be estimated from a set of $N = M+1$ time (phase) measurements spaced by the measurement interval $\tau_0$ for averaging time $\tau = m\tau_0$,

$$MDEV(\tau) = \sqrt{\frac{1}{2m^2\tau^2(N-3m+1)}\sum_{j=1}^{N-3m+1}\left\{\sum_{i=j}^{j+m-1}[x_{i+2m} - 2x_{i+m} + 2x_i]\right\}^2}. \tag{5}$$

When $m = 1$, the above equation will become the same form with the normal Allan deviation (ADEV). As an additional phase averaging operation is introduced, MDEV has the advantage of being able to distinguish between white and flicker PM noise. Moreover, the MDEV is $\sqrt{2}$ times smaller the normal ADEV, provided that the noise type of the data to be analyzed is white frequency noise[31, 37], which has been verified in Fig. 7 with simulated white noise frequency data. The importance of this conclusion is that, it is possible to use TDEV for evaluating the capability of a time transfer system in synchronizing atomic clock, despite its instability is originally specified in ADEV.

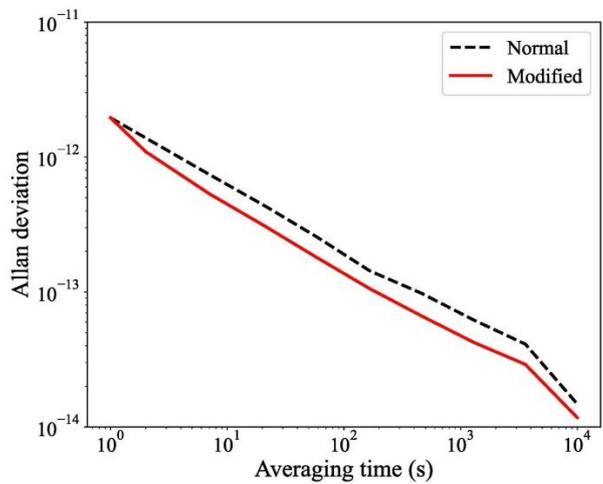

Fig. 7 Normal ADEV (black dashed line) and MDEV (red solid line) results for a set of simulated white noise frequency data.

**Funding**

National Natural Science Foundation of China (12033007, 61875205, 91836301, 12103058, and 61801458), "Western Young Scholar" Project of CAS (XAB2019B17, XAB2019B15), Frontier Science Key Research Project of CAS (QYZDB-SSW-SLH007), Strategic Priority Research Program of CAS (XDC07020200), the Youth Innovation Promotion Association, CAS (Grant No. 2022413, 2021408).